\documentclass[12pt]{article}
\newcommand{\ket}{\rangle}
\newcommand{\R}{\mbox{\boldmath $ R $}}
\newcommand{\C}{\mbox{\boldmath $ C $}}
\newcommand{\Z}{\mbox{\boldmath $ Z $}}
\newcommand{\Q}{\mbox{\boldmath $ Q $}}

\newcommand{\GCD}{\mbox{GCD} \,}
\newcommand{\LCM}{\mbox{LCM} \,}

\newcommand{\mapright}[2]
{\mathop{\hbox to 1cm{\rightarrowfill}}
\limits^{\scriptstyle #1}_{\scriptstyle #2}}

\newcommand{\mapdown}[2]
{\Big \downarrow 
\llap {$\vcenter {\hbox{$\scriptstyle #1 \,$}}$ }
\rlap {$\vcenter {\hbox{$\scriptstyle #2   $}}$ }
}

\makeatletter
\@addtoreset{equation}{section}
\makeatother

\begin{document} 
\baselineskip 7mm 
\begin{flushright}
hep-th/0205053\\
the first version: May 7, 2002
\\
revised: October 3, 2002
\end{flushright}
\vspace*{10mm}
\begin{center}
{\Large \bf 
Magnetic Translation Groups 
\vspace{1mm}\\
in an $ \mbox{\boldmath $ n $} $-dimensional Torus
\vspace{1mm}\\
and Their Representations}\footnote{%
To be published in Journal of Mathematical Physics.
}
\vspace{6mm} \\
Shogo Tanimura
\vspace*{6mm} \\
{\it
Department of Engineering Physics and Mechanics, 
Kyoto University \\ Kyoto 606-8501, Japan
}
\\
{\tt e-mail:tanimura@kues.kyoto-u.ac.jp}
\vspace{6mm}
\\
{\small Abstract}
\vspace{2mm}
\\
\begin{minipage}[t]{130mm}
{
A charged particle in a uniform magnetic field in a two-dimensional torus 
has 
a discrete noncommutative translation symmetry
instead of 
a continuous commutative translation symmetry.
We study topology and symmetry of a particle in a magnetic field
in a torus of arbitrary dimensions.
The magnetic translation group (MTG) is defined as
a group of 
translations that leave the gauge field invariant.
We show that the MTG in an $ n $-dimensional torus
is isomorphic to
a central extension of 
a cyclic group $ \Z_{\nu_1} \times \cdots \times \Z_{\nu_{2l}} \times T^m $
by $ U(1) $
with $ 2 l + m = n $.
We construct and classify irreducible unitary representations of the MTG
in a three-torus
and apply the representation theory to three examples.
We briefly describe a representation theory for a general $ n $-torus.
The MTG in an $ n $-torus can be regarded 
as a generalization of the so-called noncommutative torus.
}
\baselineskip 6mm 
\end{minipage}
\end{center}
\newpage
\baselineskip 6mm 
%
\section{Introduction}

Many people have been studying dynamics of an electrically charged particle 
in a magnetic field
for various interests.
Landau found that the energy spectrum of an electron
becomes discrete when a magnetic field is applied,
and explained the diamagnetic property of a metal.
The quantum Hall effect looked a peculiar phenomenon 
when it was first discovered
but today it is understood as a universal phenomenon
observable in a two-dimensional electron system in a magnetic field.
Dynamics of charged particles in a magnetic field 
is still an active research area.

Here we examine a group-theoretical aspect 
of the quantum system in a magnetic field.
In particular we compare symmetry in a torus 
with symmetry in a Euclidean space.
We would like to understand
how the symmetry structure of the dynamical system is affected 
by the topological structure of the underlying space.
It is known that 
the translation symmetry group becomes noncommutative 
when a uniform magnetic field is introduced into the Euclidean space.
Moreover,
the translation symmetry group becomes discrete
when the underlying space is replaced by a torus.
In this paper 
we consider a vector potential
\begin{equation}
	A 
	= 
	\sum_{j,k=1}^n x_j \omega_{jk} dx_k
	+ \sum_{j=1}^n       \alpha_j    dx_j
	\label{vector potential}
\end{equation}
over an $ n $-dimensional torus $ T^n = \R^n / \Z^n $.
Here $ \omega_{jk} $ are arbitrary integers and
$ \alpha_j $ are real numbers.
Then the corresponding magnetic field is given by the two-form
\begin{equation}
	B 
	= dA 
	= 
	\sum_{j,k=1}^n 
	\frac{1}{2} ( \omega_{jk} - \omega_{kj} ) dx_j \wedge dx_k.
	\label{curvature intro}
\end{equation}
We conclude that
the magnetic translation group (MTG) in $ T^n $ is
\begin{equation}
	S_A 
	= 
	(\R \times_\omega {\mit\Omega}^n) / (\Z \times_\omega \Z^n ),
	\label{S_A intro}
\end{equation}
where $ {\mit\Omega}^n $ is a subgroup of $ \R^n $ defined by
$
	{\mit\Omega}^n =
	\{ v \in \R^n \, | \,
	( \omega - {}^t \! \omega ) v \in \Z^n \}
$
and the group operation in $ \R \times_\omega  \R^n $ is defined by
\begin{equation}
	( x_0, x_1, \cdots, x_n ) 
	\cdot
	( y_0, y_1, \cdots, y_n ) 
	=
	( x_0 + y_0 + \sum_{j,k=1}^n x_j \omega_{jk} y_k,
	 x_1 + y_1, \cdots, x_n + y_n ).
	 \label{twist product intro}
\end{equation}
This characterization of the magnetic translation symmetry 
is one of the main results of this paper.
The MTG is actually a central extension of a cyclic group
\begin{equation}
	\Z_{\mu_1} \times \cdots \times \Z_{\mu_l} \times
	\Z_{\nu_1} \times \cdots \times \Z_{\nu_l} \times
	T^m 
	\qquad (2l+m=n)
	\label{MTG in Intro}
\end{equation}
by $ S^1 = U(1) $.
We build a complete set of irreducible representations
of the MTG in $ T^3 $.
We also describe a method to build irreducible representations of the MTG
in $ T^n $.

We would like to briefly review studies by other people 
on a quantum system in a magnetic field.
Brown \cite{Brown1964} 
found that 
the translation symmetry of an electron 
in a lattice in a uniform magnetic field
is noncommutative
and that
the quantum system obeys a projective representation of the translation group.
At the almost same time \cite{Zak1964} 
and later \cite{Zak1989}
Zak 
built a representation theory of
the lattice translation group in a magnetic field.
Ashby and Miller \cite{Ashby1965}
considered a space-time lattice of a finite size 
in uniform electric and magnetic fields
and proposed an electromagnetic translation group.
Avron, Herbst, and Simon have been studying
spectral problems of the Schr{\"o}dinger operators in a magnetic field
in a series of papers
\cite{Avron1978-1}-\cite{Avron1985}.
Particularly, in \cite{Avron1978-2}
they examined a system of particles in a uniform magnetic field 
and characterized a constant of motion analogous to the total momentum.
Dubrovin and Novikov \cite{Dubrovin1980, Novikov1981}
studied the spectrum of the Pauli operator 
in a two-dimensional lattice with a periodic magnetic field
and intensively analyzed the gap structure above the ground state.
Asch, Over, and Seiler \cite{Asch1994}
clarified 
how the inequivalent Hamiltonians on a torus in a magnetic field
are induced from
a Hamiltonian on the universal covering space of the torus.
In a series of studies \cite{Lulek1994}-\cite{Florek1996-2}
Lulek, Florek, Lipinski, and Walcerz
established a systematic method to construct 
central extensions of a finitely generated Abelian group.
Their results are equivalent to the MTGs in a lattice.
Kuwabara \cite{Kuwabara1995, Kuwabara1999} is studying 
relations between 
the trajectories of a classical particle 
and the spectra of its quantized system
and has obtained many results.
Gruber \cite{Gruber} also examined
quantization of a particle on a Riemannian manifold in a magnetic field
from a view point of geometric quantization.

As reviewed above 
a lot of studies on dynamics and symmetry in a magnetic field
have been done.
Although 
MTGs in a finite lattice and in an infinite lattice 
have been much investigated,
the MTG in a torus of arbitrary dimensions is not yet fully investigated.
Motivated by a recent study \cite{Antoniadis, Arkani-Hamed}
on extra dimensions of the space-time,
Sakamoto et al. \cite{Sakamoto1999, Sakamoto2000, Sakamoto2001}
are developing field theoretical models 
in which the translation symmetry of an extra circle is 
spontaneously broken by a nontrivial boundary condition in the extra $ S^1 $.
Moreover, we are developing models
\cite{Tanimura2001,Tanimura2002}
in which the rotation symmetry of an extra two-sphere is 
spontaneously broken by a magnetic monopole in the extra $ S^2 $.
So we would like to understand 
how a background gauge field in a compact space influences
symmetry structure of a quantum system.
Hence we decide to investigate symmetry
in a magnetic field in a torus.

This paper is organized as follows.
In Sec. 2 we shall examine 
how symmetry of a quantum system in a magnetic field is changed 
when the underlying two-dimensional Euclidean space is replaced by
a two-dimensional torus.
In Sec. 3 we extend our discussion to an $ n $-dimensional torus.
We introduce a noncommutative group structure into $ \R^{n+1} $
and use it to construct a magnetic fiber bundle,
which is a bundle over $ T^n $ with a fiber $ S^1 $.
In Sec. 4 we classify topological structures of the bundles.
In Sec. 5 we define connections, 
which are generalizations of a vector potential,
and classify them.
In Sec. 6 we define a magnetic translation group
as a group of lifted translations that leave the connection invariant.
In Sec. 7 we build a representation theory of the MTG for $ T^3 $
and illustrate the theory by a few examples.
In Sec. 8 we describe an outline of the representation theory of the MTG
for a general $ T^n $.
Section 9 is devoted to conclusions and discussions.
To reach the main result quickly
the reader may read only sections 3, 5, and 6.

\section{Symmetries in a magnetic field}
This section is devoted to 
exercises to get ideas about the problem.
The reader may skip this section and restart from Sec. 3 
without missing the main course of the paper.

\subsection{Euclidean space}
Let us begin our discussion by examining symmetry
of quantum mechanics of a particle 
in the uniform magnetic field in $ \R^2 $.
It is a well-known system 
and becomes a starting point to explore further nontrivial systems.

A uniform magnetic field
$ B dx \wedge dy = dA $
is derived from a vector potential
$ A = B x \, dy $.
The Schr{\"o}dinger equation 
is
\begin{equation}
	H \psi =
	\left[
		- \frac12 \bigg( \frac{\partial}{\partial x}       \bigg)^2
		- \frac12 \bigg( \frac{\partial}{\partial y} - iBx \bigg)^2
	\right] 
	\psi(x,y)
	= E \psi.
	\label{Sch}
\end{equation}
Then the operators 
\begin{equation}
	\tilde{P}_x := -i \frac{\partial}{\partial x} - By,
	\qquad
	P_y := -i \frac{\partial}{\partial y}
	\label{Sch cons}
\end{equation}
commute with $ H $.
These generate unitary transformations
\begin{eqnarray}
	&&
	( U_x(a) \psi ) (x,y)
	= e^{-i \tilde{P}_x a} \psi (x,y)
	= e^{i B a y} \psi(x-a,y),
	\label{U_x}\\
	&&
	( U_y(b) \psi ) (x,y)
	= e^{-i P_y b} \psi (x,y)
	= \psi(x,y-b).
	\label{U_y}
\end{eqnarray}
It is to be noted that
$ U_x(a) $ is a combination of a translation in the $ x $-direction
by the length $ a $
and a gauge transformation.
It is also to be noted that
the translation in the $ x $-direction and the one in the $ y $-direction 
do not commute but satisfy
\begin{equation}
	  U_x(a) 
	  U_y(b) 
	( U_x(a) )^{-1} 
	( U_y(b) )^{-1} 
	= e^{iBab}.
	\label{noncom}
\end{equation}

The momentum generates a continuous symmetry
and enables us to separate the variables.
For example, if we put the eigenvalue of $ P_y $ as $ k $,
the wave function is factorized as
\begin{equation}
	\psi(x,y) = e^{iky} \phi(x).
\end{equation}
Then the Schr{\"o}dinger equation (\ref{Sch}) is rewritten as
\begin{equation}
	H \psi =
	e^{iky}
	\left[
		- \frac12 \left( \frac{\partial}{\partial x} \right)^2
		+ \frac12 \left( k - Bx \right)^2
	\right] 
	\phi(x)
	= e^{iky} E \phi(x)
\end{equation}
and is reduced to the equation of a harmonic oscillator.
Hence the energy eigenvalues are given by
\begin{equation}
	E = |B| 
	\left( n + \frac{1}{2} \right) 
	\qquad (n=0,1,2,\cdots)
	\label{E}
\end{equation}
and are called the Landau levels.
Each eigenvalue is infinitely degenerated with respect to 
$ - \infty < k < \infty $.

\subsection{Torus}

Next we turn to a two-dimensional torus.
The two-torus $ T^2 $
is defined as the quotient space $ \R^2 / \Z^2 $.
Namely, the points in $ \R^2 $
\begin{equation}
	(x,y) \sim
	(x+1,y) \sim
	(x,y+1) 
\end{equation}
are identified as a single point in $ T^2 $.
If we impose a pseudoperiodic condition
\begin{equation}
	\psi(x+1,y) = e^{iBy} \psi(x,y),
	\qquad
	\psi(x,y+1) = \psi(x,y)
	\label{pc}
\end{equation}
on the wave function,
the Schr{\"o}dinger equation (\ref{Sch}) is well defined over $ T^2 $.
In other words, on the space of functions satisfying 
the pseudoperiodic condition,
the operator $ H $ becomes self-adjoint.
To make the two conditions in (\ref{pc}) compatible each other
we need to have
\begin{eqnarray}
	\psi(x+1,y+1) 
& = & 	e^{iB(y+1)} \psi(x,y+1) 
	= e^{iB} e^{iBy} \psi(x,y)
	\nonumber \\
& = &	\psi(x+1,y) = e^{iBy} \psi(x,y).
\end{eqnarray}
Hence we should have $ e^{iB} = 1 $.
Namely, in the magnetic field strength
\begin{equation}
	B = 2 \pi \nu
	\label{charge}
\end{equation}
$ \nu $ must be an integer.
We call $ \nu $ the magnetic flux number of the torus.

The operators $ \tilde{P}_x $ and $ P_y $ in (\ref{Sch cons})
commute with $ H $ defined in (\ref{Sch}).
However, 
when they act on a wave function satisfying 
the pseudoperiodic condition (\ref{pc}),
they do not give back 
a function satisfying the pseudoperiodic condition 
but instead give
\begin{eqnarray}
	P_y \psi(x+1,y) & = & e^{iBy} (P_y + B) \psi(x,y), 
	\\
	\tilde{P}_x \psi(x,y+1) & = & ( \tilde{P}_x - B ) \psi(x,y).
\end{eqnarray}
Hence, the actions of these operators are not closed 
in the space of pseudoperiodic functions.
Thus we get a lesson that
{\it the generator of infinitesimal translation does not exist in the torus.}
However,
it is still possible to construct operators for finite translations.
We let the finite translation operators (\ref{U_x}) and (\ref{U_y}) act
on a pseudoperiodic function (\ref{pc}),
and examine whether the resultant functions satisfy 
the pseudoperiodic condition.
Using the flux quantization (\ref{charge}) we get
\begin{eqnarray}
	( U_x(a) \psi ) (x,y+1)
	& = & e^{i B a (y+1)} \psi(x-a,y+1) \nonumber \\
	& = & e^{i B a} e^{i B a y} \psi(x-a,y) \nonumber \\
	& = & e^{2 \pi i \nu a} ( U_x(a) \psi ) (x,y),
	\\
	( U_y(b) \psi ) (x+1,y)
	& = & \psi(x+1,y-b) \nonumber \\
	& = & e^{i B (y-b)} \psi(x,y-b) \nonumber \\
	& = & e^{-i B b} e^{i B y} \psi(x,y-b) \nonumber \\
	& = & e^{- 2 \pi i \nu b} e^{i B y} ( U_y(b) \psi ) (x,y).
\end{eqnarray}
Therefore,
the transformed wave functions,
$ U_x (a) \psi $ and $ U_y (b) \psi $, satisfy
the pseudoperiodic condition (\ref{pc}) if and only if
\begin{equation}
	\nu a, \nu b \in \Z.
	\label{discrete}
\end{equation}
Consequently,
the lengths of shifts, $ a $ and $ b $,
are restricted to integral multiples of $ 1/\nu $.
Moreover, on a pseudoperiodic function the shifts 
by the unit length act as
\begin{eqnarray}
	&&
	(U_x (1) \psi) (x,y)
	= e^{iBy} \psi(x-1,y) 
	= \psi(x,y),
	\\ &&
	(U_y (1) \psi) (x,y)
	= \psi(x,y-1) 
	= \psi(x,y).
\end{eqnarray}
Hence $ U_x(1) $ and $ U_y(1) $ are identity operators.
Thus the operators $ U_x(1/\nu) $ and $ U_y(1/\nu) $ generate
a cyclic group $ \Z_\nu = \Z / \nu \Z $ of the order $ \nu $.
However, as seen in (\ref{noncom}) 
their commutator produces a nontrivial phase factor.
Thus we conclude that
the symmetry of the quantum system in the torus magnetic field
is described by a projective representation of $ \Z_\nu \times \Z_\nu $.

The group of translations of the quantum system in the magnetic field 
is called a magnetic translation group (abbreviated as MTG).
A more precise definition of the MTG will be given in Sec. 6.
In the torus the MTG becomes discrete and finite.
Its representation is constructed as follows.
Let 
$ \{ | 0 \ket, \, | 1 \ket, \cdots, | \nu - 1 \ket \} $
be a basis of the representation space.
Then we define the action of the translation operators by
\begin{eqnarray}
	U_x ( n_x / \nu ) | q \ket 
	& = & 
	e^{ 2 \pi i n_x q/\nu} | q \ket,
	\label{rep Ux}
	\\
	U_y ( n_y / \nu ) | q \ket 
	& = & 
	| q + n_y \, (\mbox{mod} \, \nu ) \ket,
	\label{rep Uy}
\end{eqnarray}
for $ n_x, n_y \in \Z $.
We can easily verify that they satisfy
\begin{equation}
	U_x ( n_x / \nu ) 
	U_y ( n_y / \nu ) 
	\left( U_x ( n_x / \nu ) \right)^{-1} 
	\left( U_y ( n_y / \nu ) \right)^{-1} 
	| q \ket
	\nonumber \\
= 
	e^{ i ( 2 \pi \nu) (n_x/\nu) (n_y /\nu)} 
	| q \ket,
\end{equation}
which is homomorphic to the commutator (\ref{noncom}).
This representation is irreducible and its dimension is $ \nu $.
Hence each energy eigenvalue (\ref{E}) is degenerated by $ \nu $ folds.

\subsection{Three-torus}
Let us examine 
the case of a three-dimensional torus briefly
to motivate further discussion.
With real constants $ ( b_1, b_2, b_3 ) $
a vector potential
\begin{equation}
	A 
	= b_1 x_2 d x_3
	+ b_2 x_3 d x_1
	+ b_3 x_1 d x_2
	\label{A 3-torus}
\end{equation}
gives rise to a magnetic field
\begin{equation}
	B = dA 
	= b_1 d x_2 \wedge d x_3
	+ b_2 d x_3 \wedge d x_1
	+ b_3 d x_1 \wedge d x_2.
	\label{B 3-torus}
\end{equation}
The Hamiltonian is then given by
\begin{equation}
	H \psi =
	- \frac12 
	\left[
	  \bigg( \frac{\partial}{\partial x_1} - i b_2 x_3 \bigg)^2
	+ \bigg( \frac{\partial}{\partial x_2} - i b_3 x_1 \bigg)^2
	+ \bigg( \frac{\partial}{\partial x_3} - i b_1 x_2 \bigg)^2
	\right] 
	\psi(x_1,x_2,x_3).
	\label{Sch on 3-torus}
\end{equation}
On the three-torus the wave function must satisfy 
a set of conditions
\begin{eqnarray}
	&&
	\psi(x_1+1,x_2,x_3) = e^{i b_3 x_2} \psi(x_1,x_2,x_3),
	\nonumber \\
	&&
	\psi(x_1,x_2+1,x_3) = e^{i b_1 x_3} \psi(x_1,x_2,x_3),
	\nonumber \\
	&&
	\psi(x_1,x_2,x_3+1) = e^{i b_2 x_1} \psi(x_1,x_2,x_3),
	\label{pc on 3-torus}
\end{eqnarray}
which is a generalization of the 
the pseudoperiodic condition (\ref{pc}) of the two-torus.

We would like to find 
a complete set of translation operators
that 
commute with $ H $ and 
are compatible with the pseudoperiodic condition (\ref{pc on 3-torus}).
Of course, if the magnetic field is parallel to one of the axes,
the system is reduced to the two-torus
as has been discussed by Zak \cite{Zak1989}.
For example, if $ ( b_1, b_2, b_3 ) = ( 0, 0, B ) $,
the Hamiltonian (\ref{Sch on 3-torus}) and
the condition (\ref{pc on 3-torus}) are reduced
to (\ref{Sch}) and (\ref{pc}), respectively.
However, it is a highly nontrivial 
and not yet fully solved problem
to find a complete symmetry group for an inclined magnetic field
$ ( b_1, b_2, b_3 ) $.
Thus we decide to develop a more systematic method
to construct the translation symmetry group for a generic magnetic field
in the $ n $-torus.

\section{Magnetic fiber bundle}

We shall extend the previous consideration on the two-dimensional torus
to arbitrary dimensions.
What we will do in the rest of this paper is
to construct $ U(1) $ principal fiber bundles over 
an $ n $-dimensional torus $ T^n $,
to classify the bundles,
to introduce
$ U(1) $ connections with constant curvatures over $ T^n $,
to define the MTG as the stability group of each connection,
and to construct the representations of the MTGs.
Throughout this paper we are identifying $ S^1 $ with $ U(1) $.

Let us begin with 
construction of $ S^1 $ principal fiber bundles over $ T^n $.
For this purpose we introduce a noncommutative group structure into $ \R^{n+1} $
as follows.
Take an $ n \times n $ matrix $ \omega $ which consists of integers, 
$ \omega_{jk} \in \Z $ $ (j,k=1,\cdots,n) $.
The matrix $ \omega $ is not necessarily antisymmetric. 
Define a product of 
$ ( x_0, x_1, \cdots, x_n ), ( y_0, y_1, \cdots, y_n ) \in \R^{n+1} $
by
\begin{equation}
	( x_0, x_1, \cdots, x_n ) 
	\cdot
	( y_0, y_1, \cdots, y_n ) 
	:=
	( x_0 + y_0 + \sum_{j,k=1}^n x_j \omega_{jk} y_k,
	 x_1 + y_1, \cdots, x_n + y_n ).
	 \label{twist product}
\end{equation}
In the following
we abbreviate the notation of the vectors as 
$ x = (x_1, \cdots, x_n) \in \R^n $.
We write the inner product of vectors as
$ x y = \sum_{j=1}^n x_j \, y_j $ 
and the bilinear form as
$ x \omega y = \sum_{j,k=1}^n x_j \omega_{jk} y_k $.
It is easily verified that the set $ \R^{n+1} $ becomes a group 
with this product operation;
the associativity is satisfied as
\begin{eqnarray}
	( ( x_0, x ) \cdot ( y_0, y ) ) \cdot ( z_0, z ) 
& = &
	( x_0 + y_0 + x \omega y, x+y ) \cdot ( z_0, z ) 
	\nonumber \\
& = &
	( x_0 + y_0 + z_0 + x \omega y + (x+y) \omega z, (x+y)+z )
	\nonumber \\
& = &
	( x_0 + y_0 + z_0 + x \omega y + x \omega z + y \omega z, x+y+z )
	\nonumber \\
& = &
	( x_0 + y_0 + z_0 + x \omega (y+z) + y \omega z, x+(y+z) )
	\nonumber \\
& = &
	( x_0, x ) \cdot ( ( y_0, y ) \cdot ( z_0, z ) ),
	 \label{associativity}
\end{eqnarray}
the unit element is given by $ (0,0) \in \R \times \R^n $,
and the inverse element of $ (x_0, x) \in \R \times \R^n $ is given by
\begin{equation}
	( x_0, x )^{-1} 
	= ( - x_0 + x \omega x, -x ).
	 \label{inverse}
\end{equation}
The set $ \R^{n+1} $ equipped with this group structure
is denoted by $ \R \times_\omega \R^n $.
A commutator is calculated as
\begin{eqnarray}
&&
	( x_0, x ) \cdot ( y_0, y ) \cdot ( x_0, x )^{-1} \cdot ( y_0, y )^{-1}
	\nonumber \\
& = &
	( x_0 + y_0 + x \omega y, x+y ) 
	\cdot
	( - x_0 + x \omega x, -x )
	\cdot
	( - y_0 + y \omega y, -y )
	\nonumber \\
& = &
	( y_0 + x \omega y + x \omega x - (x+y) \omega x, y ) 
	\cdot
	( - y_0 + y \omega y, -y )
	\nonumber \\
& = &
	( x \omega y + x \omega x - (x+y) \omega x 
	+ y \omega y - y \omega y, 0 ) 
	\nonumber \\
& = &
	( x \omega y - y \omega x, 0 ),
	 \label{commutativity}
\end{eqnarray}
and therefore $ \R \times_\omega \R^n $ is Abelian 
if and only if $ \omega $ is a symmetric matrix.
The natural projection map $ \R \times_\omega \R^n \to \R^n $ 
becomes a group homomorphism.
As its kernel $ \R \times_\omega \{0\} $ is contained in the center of
$ \R \times_\omega \R^n $,
the group $ \R \times_\omega \R^n $ is a central extension 
of $ \R^n $ by $ \R $.

The subset 
$ \Z \times_\omega \Z^n = \{ (m_0, m_1, \cdots, m_n) \, | \, 
m_0, m_j \in \Z \} $ 
is also a subgroup of $ \R \times_\omega \R^n $
but it is not isomorphic to 
the standard Abelian group $ \Z^{n+1} $.
The subgroup 
$ \Z \times_\omega \Z^n $ acts freely on 
$ \R \times_\omega \R^n $ from the left
via the group operation.
Hence the space of orbits 
\begin{equation}
	P^{n+1}_\omega 
	:= 
	(\Z \times_\omega \Z^n) \backslash (\R \times_\omega \R^n) 
	\label{P}
\end{equation}
becomes a smooth manifold.

The group operation also induces
action of the group $ \R \times_\omega \R^n $ on the space $ P^{n+1}_\omega $
from the right.
The subgroups
$ \Z \times_\omega \{ 0 \} \subset
  \R \times_\omega \{ 0 \} $
are contained in the center of 
$ \R \times_\omega \R^n $
and hence their actions from the right are equivalent to those from the left.
The subgroups 
$ \R \times_\omega \{ 0 \} $ and 
$ \Z \times_\omega \{ 0 \} $ are
isomorphic to $ \R $ and $ \Z $, respectively.
Thus 
$ \R $ acts on $ P^{n+1}_\omega $
but its subgroup $ \Z $ acts trivially on $ P^{n+1}_\omega $
since $ \Z $ is contained in the dividing group $ \Z \times_\omega \Z^n $
of the quotient space (\ref{P}).
Therefore the action of $ \R $ is reduced to
the effective action of $ S^1 = \R/\Z $ on $ P^{n+1}_\omega $.
The space of orbit $ P^{n+1}_\omega/S^1 $ is diffeomorphic to a torus $ T^n $.
Consequently we obtain a principal fiber bundle 
with the canonical projection map $ \pi_\omega : P^{n+1}_\omega \to T^n $
with a structure group $ S^1 $.
We call this fiber bundle 
a {\it magnetic fiber bundle twisted by the matrix} $ \omega $.
The procedure to construct the magnetic fiber bundle is summarized by
the following commutative diagram
\begin{eqnarray}
   \begin{array}{ccccc}
      \Z \times_\omega \{0\} & \to & \Z \times_\omega \Z^n & \to & \Z^n \\
      \downarrow             &     & \downarrow            &     & \downarrow \\
      \R \times_\omega \{0\} & \to & \R \times_\omega \R^n & \to & \R^n \\
      \downarrow             &     & \downarrow            &     & \downarrow \\
      S^1                    & \to & P^{n+1}_\omega        & 
      \mapright{}{\pi_\omega} & T^n 
   \end{array}
\end{eqnarray}

A function 
$ f :  P^{n+1}_\omega \to \C $ is identified with a function 
$ f : \R \times_\omega \R^n \to \C $ 
that is invariant under action of $ \Z \times_\omega \Z^n $ from the left as
\begin{equation}
	f (m_0 + x_0 + m \omega x, m + x )
	=
	f (x_0, x ),
	\qquad
	(m_0, m) \in \Z \times_\omega \Z^n.
	\label{invariance}
\end{equation}
Moreover, when the function $ f : P^{n+1}_\omega \to \C $ satisfies
\begin{equation}
	f (x_0 + t, x )
	=
	e^{-2 \pi i t} \, f (x_0, x ),
	\qquad
	t \in \R,
	\label{equivariance}
\end{equation}
it is called an equivariant function on $ P^{n+1}_\omega $.
Hence the equivariant function $ f $ has the property
\begin{equation}
	f (x_0, x + m )
	=
	e^{ 2 \pi i m \omega x } \, f (x_0, x ),
	\qquad
	m \in \Z^n.
	\label{twist bc}
\end{equation}
This is a generalization of the pseudoperiodic condition (\ref{pc})
\begin{equation}
	\psi(x+1,y) = e^{2 \pi i \nu y} \psi(x,y), \qquad
	\psi(x,y+1) = \psi(x,y).
	\label{twist bc2}
\end{equation}
In fact, if we take the matrix
\begin{equation}
	\omega =
	\left(
		\begin{array}{rr}
		0 & \nu \\
		0 &  0 
		\end{array}
	\right),
	\label{T^2}
\end{equation}
the general condition (\ref{twist bc}) of $ T^n $
is reduced to the specific one (\ref{twist bc2}) of $ T^2 $.

\section{Equivalent magnetic bundles}

In the above construction
each magnetic fiber bundle is specified by an integral matrix $ \omega $.
However, it can happen that different matrices $ \omega $ and $ \omega' $
give rise to equivalent fiber bundles.
In this section we prove that
$ \omega $ and $ \omega' $ induce equivalent fiber bundles
if and only if
the difference $ \omega' - \omega $ is a symmetric integral matrix.
Therefore, we may choose a representative matrix $ \omega $
such that $ \omega_{jk} = 0 $ for $ j \ge k $.
Namely, 
the upper triangle matrix
\begin{equation}
	\omega =
	\left(
		\begin{array}{llllll}
		0 & \omega_{12} & \omega_{13} & 
		\cdots & \omega_{1,n-1} & \omega_{1n} \\
		0 & 0           & \omega_{23} & 
		\cdots & \omega_{2,n-1} & \omega_{2n} \\
		0 & 0           & 0           & 
		\cdots & \omega_{3,n-1} & \omega_{3n} \\
		\vdots & \vdots & \vdots & \ddots & \vdots & \vdots \\
		0 & 0           & 0           & 
		\cdots & 0 & \omega_{n-1,n} \\
		0 & 0 & 0 & \cdots & 0 & 0
		\end{array}
	\right)
	\label{standard omega}
\end{equation}
with integers $ \omega_{jk} $
can be taken as a standard form of the matrix $ \omega $.
The reader will not miss the main result of the paper
even if he skips this section and restarts from Sec. 5 

Here we introduce three kinds of isomorphisms that convert
a bundle specified by a matrix $ \omega $
to a bundle specified by another matrix $ \omega' $.

Let us introduce the first kind of bundle isomorphism.
When a symmetric matrix $ \sigma $ of integral elements, 
$ \sigma_{jk} = \sigma_{kj} \in \Z $,
satisfies
\begin{equation}
	\sum_{j,k=1}^n m_j \sigma_{jk} m_k \in 2 \Z
	\label{integral symmetric form}
\end{equation}
for any $ m = (m_1, \cdots, m_n) \in \Z^n $,
we call $ \sigma $ an even symmetric matrix.
This requirement for $ \sigma $ is equivalent to demanding that
the off-diagonal elements $ \sigma_{jk} $ are integers and that
the diagonal elements $ \sigma_{jj} $ are even integers.
Here we will show that
two magnetic bundles $ P^{n+1}_\omega $ and $ P^{n+1}_{\omega+\sigma} $
are isomorphic each other for any even symmetric matrix $ \sigma $.
For this purpose let us define a map 
$ \phi_\sigma : \R \times_\omega \R^n \to \R \times_{\omega+\sigma} \R^n $
by
\begin{equation}
	\phi_\sigma ( x_0, x ) := ( x_0 + \frac{1}{2} x \sigma x, x ).
	\label{iso map}
\end{equation}
Existence of the inverse map is obvious; it is given by
$ \phi_\sigma^{-1} ( x_0, x ) = ( x_0 - \frac{1}{2} x \sigma x, x ) $.
It is easily verified that the map $ \phi_\sigma $ is a group isomorphism as
\begin{eqnarray}
	\phi_\sigma ( ( x_0, x ) \cdot_\omega ( y_0, y ) )
& = &
	\phi_\sigma ( x_0 + y_0 + x \omega y, x+y )
	\nonumber \\
& = &
	( x_0 + y_0 
	+ x \omega y + \frac{1}{2} (x+y) \sigma (x+y), x+y )
	\nonumber \\
& = &
	( x_0 + y_0 
	+ \frac{1}{2} x \sigma x + \frac{1}{2} y \sigma y + x (\omega+\sigma) y,
	x+y )
	\nonumber \\
& = &
	( x_0 + \frac{1}{2} x \sigma x, x ) \cdot_{\omega+\sigma}
	( y_0 + \frac{1}{2} y \sigma y, y ) 
	\nonumber \\
& = &
	\phi_\sigma ( x_0, x ) \cdot_{\omega+\sigma}
	\phi_\sigma ( y_0, y ),
	\label{homomorphism}
\end{eqnarray}
where we have distinguished the product operation of 
$ \R \times_{\omega+\sigma} \R^n $ from that of $ \R \times_\omega \R^n $.
The map $ \phi_\sigma $ sends the integer subgroup 
$ \Z \times_\omega \Z^n $ to 
$ \Z \times_{\omega+\sigma} \Z^n $, 
since $ \sigma $ is even as required in (\ref{integral symmetric form}).
Therefore, $ \phi_\sigma $ induces a diffeomorphism
\begin{equation}
	(\phi_\sigma)_* :
	( \Z \times_\omega \Z^n ) 
	\backslash 
	( \R \times_\omega \R^n ) 
	\to
	( \Z \times_{\omega+\sigma} \Z^n ) 
	\backslash 
	( \R \times_{\omega+\sigma} \R^n ).
	\label{bundle map}
\end{equation}
Moreover, since
$ \phi_\sigma $ is the identity map when it is restricted on 
$ \R \times_\omega \{0\} $,
\begin{eqnarray}
	\phi_\sigma ( (t, 0) \cdot_\omega (x_0, x)  )
& = &
	\phi_\sigma (t, 0) \cdot_{\omega + \sigma} \phi_\sigma ( x_0, x )
	\nonumber \\
& = &
	(t, 0) \cdot_{\omega + \sigma} \phi_\sigma ( x_0, x ),
	\label{equivariance of phi_sigma}
\end{eqnarray}
thus $ (\phi_\sigma)_* $ is equivariant with respect to
the action of $ S^1 $.
It is also clear that $ \pi_\omega = \pi_{\omega+\sigma} \circ (\phi_\sigma)_* $.
Thus we conclude that the map $ (\phi_\sigma)_* $ is
an isomorphism between the principal fiber bundles
$ P^{n+1}_\omega $ and $ P^{n+1}_{\omega+\sigma} $.

Next we shall introduce the second kind of bundle isomorphism.
We identify a diagonal matrix 
$ \Delta = \mbox{diag} ( \Delta_1, \Delta_2, \cdots, \Delta_n ) $
with a vector
$ \Delta = $ $ ( \Delta_1, \Delta_2, \cdots, \Delta_n ) $ $ \in \Z^n $.
Then we define a map
$ \phi_\Delta : \R \times_\omega \R^n \to \R \times_{\omega+\Delta} \R^n $
by
\begin{equation}
	\phi_\Delta ( x_0, x ) 
	:= 
	( x_0 + \frac{1}{2} x \Delta x + \frac{1}{2} \Delta x, x )
	=
	( x_0 
	+ \frac{1}{2} \sum_{j=1}^n ( x_j \Delta_j x_j + \Delta_j x_j ), x ).
	\label{iso map odd}
\end{equation}
It is also easily verified that 
$ \phi_\Delta $ is a group isomorphism as
\begin{eqnarray}
	\phi_\Delta ( ( x_0, x ) \cdot_\omega ( y_0, y ) )
& = &
	\phi_\Delta ( x_0 + y_0 + x \omega y, x+y )
	\nonumber \\
& = &
	( x_0 + y_0 
	+ x \omega y 
	+ \frac{1}{2} (x+y) \Delta (x+y) + \frac{1}{2} \Delta (x+y),
	x+y )
	\nonumber \\
& = &
	( x_0 + y_0 
	+ \frac{1}{2} x \Delta x + \frac{1}{2} \Delta x
	+ \frac{1}{2} y \Delta y + \frac{1}{2} \Delta y
	+ x (\omega+\Delta) y,
	x+y )
	\nonumber \\
& = &
	( x_0 
	+ \frac{1}{2} x \Delta x + \frac{1}{2} \Delta x,
	x )
	\cdot_{\omega+\Delta}
	( y_0 
	+ \frac{1}{2} y \Delta y + \frac{1}{2} \Delta y,
	y ) 
	\nonumber \\
& = &
	\phi_\Delta ( x_0, x ) \cdot_{\omega+\Delta}
	\phi_\Delta ( y_0, y ).
	\label{Delta homomorphism}
\end{eqnarray}
Note that when $ x_j $ is an integer,
$ x_j^2 + x_j = x_j ( x_j + 1 ) $ is always an even integer
and hence
$ \frac{1}{2} \Delta_j (x_j^2 + x_j) $ is an integer.
Therefore the map 
$ \phi_\Delta $ sends the integer subgroup 
$ \Z \times_\omega \Z^n $ to 
$ \Z \times_{\omega+\Delta} \Z^n $.
Moreover,
$ \phi_\Delta $ sends 
$ \R \times_\omega \{0\} $ to
$ \R \times_{\omega + \Delta} \{0\} $
identically.
Thus
the induced map $ (\phi_\Delta)_* $ becomes
an isomorphism between the principal fiber bundles
$ P^{n+1}_\omega $ and $ P^{n+1}_{\omega+\Delta} $.

There is the third kind of bundle isomorphism, 
which will be used when we classify connections later.
For each
$ \varepsilon = $ 
$ ( \varepsilon_1, \varepsilon_2, \cdots, \varepsilon_n ) $ $ \in \Z^n $
we define a map
$ \phi_\varepsilon : \R \times_\omega \R^n \to \R \times_{\omega} \R^n $
by
\begin{equation}
	\phi_\varepsilon ( x_0, x ) 
	:= 
	( x_0 + \varepsilon x, x )
	=
	( x_0 + \sum_{j=1}^n \varepsilon_j x_j, x ).
	\label{iso map eps}
\end{equation}
It is also easily verified that 
$ \phi_\varepsilon $ is a group isomorphism as
\begin{eqnarray}
	\phi_\varepsilon ( ( x_0, x ) \cdot_\omega ( y_0, y ) )
& = &
	\phi_\varepsilon ( x_0 + y_0 + x \omega y, x+y )
	\nonumber \\
& = &
	( x_0 + y_0 
	+ x \omega y 
	+ \varepsilon (x+y),
	x+y )
	\nonumber \\
& = &
	( x_0 
	+ \varepsilon x
	+ y_0 
	+ \varepsilon y
	+ x \omega y,
	x+y )
	\nonumber \\
& = &
	( x_0 
	+ \varepsilon x,
	x )
	\cdot_{\omega}
	( y_0 
	+ \varepsilon y,
	y ) 
	\nonumber \\
& = &
	\phi_\varepsilon ( x_0, x ) \cdot_{\omega}
	\phi_\varepsilon ( y_0, y ).
	\label{eps homomorphism}
\end{eqnarray}
The map 
$ \phi_\varepsilon $ sends the integer subgroup 
$ \Z \times_\omega \Z^n $ to 
$ \Z \times_\omega \Z^n $.
Moreover,
$ \phi_\varepsilon $ sends 
$ \R \times_\omega \{0\} $ to
$ \R \times_\omega \{0\} $
identically.
Thus
the group isomorphism $ \phi_\varepsilon $ induces
an automorphism 
$ (\phi_\varepsilon)_* $ of the principal fiber bundle $ P^{n+1}_\omega $.

As a summary, we write down a combined isomorphism of the three kinds of maps
\begin{equation}
	(     \phi_\varepsilon 
	\circ \phi_\Delta 
	\circ \phi_\sigma )
	( x_0, x ) := 
	( x_0 
	+ \frac{1}{2} x (\sigma + \Delta) x
	+ \frac{1}{2} \Delta x
	+ \varepsilon x, x ).
	\label{combined map}
\end{equation}
By adding 
an integral diagonal matrix $ \Delta $
to an even symmetric matrix $ \sigma $, 
we can make any integral symmetric matrix $ \sigma' = \sigma + \Delta $.
Therefore,
by combining the first and second kinds of isomorphisms,
$ \phi_\sigma $ and $ \phi_\Delta $,
we can establish
an isomorphism between $ P^{n+1}_{\omega} $ and $ P^{n+1}_{\omega + \sigma'} $
for any integral symmetric matrix $ \sigma' $.
In other words, the set of magnetic fiber bundles has 
a one-to-one correspondence with
$ \mbox{Mat} (n, \Z) / \mbox{Sym} (n, \Z) $,
where the quotient is taken in the sense of additive groups.

\section{Connection}

In this section we define the vector potentials
that yield
uniform magnetic fields in an $ n $-dimensional torus.
We use the words, a vector potential, a gauge field, and a connection,
to describe the same notion.
Magnetic field strength and curvature are an identical notion.

Let us define a differential one-form $ A $ on $ \R \times_\omega \R^n $ by
\begin{equation}
	A 
	:= 
	- dx_0
	+ \sum_{j,k=1}^n x_j \omega_{jk} dx_k
	+ \sum_{j=1}^n       \alpha_j    dx_j
	= 
	- dx_0 + x \omega dx + \alpha dx
	\label{connection}
\end{equation}
with a real vector $ \alpha \in \R^n $.
These parameters $ \alpha = (\alpha_1, \cdots, \alpha_n) $ characterize
the Aharonov-Bohm effect.
The action of 
$ (m_0,m) \in \Z \times_\omega \Z^n $ from the left of 
$ \R \times_\omega \R^n $ defines a map
$ \varphi : (x_0,x) \mapsto (m_0 + x_0 + m \omega x, m + x) $.
Note that the one-form $ A $
is invariant under the transformation by $ \varphi $ as
\begin{equation}
	\varphi^* A 
	= 
	- ( dx_0 + m \omega dx )
	+ ( m+x ) \omega dx
	+         \alpha dx
	= A.
	\label{invariance of connection}
\end{equation}
Thus $ A $ can be regarded as a one-form on 
$ P^{n+1}_\omega = 
( \Z \times_\omega \Z^n ) \backslash ( \R \times_\omega \R^n ) $.
It is also obvious that $ A $ is invariant under 
a transformation $ (x_0, x) \mapsto (x_0+t, x) $ for any $ t \in \R $.
Moreover, $ A $ satisfies
\begin{equation}
	\left< \frac{\partial}{\partial x_0}, A \right> = -1
	\label{vertical value of A}
\end{equation}
by the definition.
In the above equation,
$ \left< \cdot , \cdot \right> $ 
denotes the pairing of a vector and a one-form.
Thus
$ A $ satisfies the axiom of a connection form
of the principal bundle $ \pi_\omega : P^{n+1}_\omega \to T^n $.

We can classify the connections using
isomorphism maps introduced in the last section.
The connection $ A_{\omega,\alpha} $ defined by (\ref{connection})
is parametrized 
by an integral matrix $ \omega \in \mbox{Mat} (n,\Z) $ 
and a real vector $ \alpha \in \R^n $.
For any 
even symmetric matrix $ \sigma \in \mbox{EvenSym} (n,\Z) $ 
and integral vectors $ \Delta, \varepsilon \in \Z^n $,
the combined isomorphism (\ref{combined map}) induces a transformation
\begin{eqnarray}
&&
	(     \phi_\varepsilon 
	\circ \phi_\Delta 
	\circ \phi_\sigma )^*
	A_{\omega+\sigma+\Delta, \alpha+\frac{1}{2} \Delta + \varepsilon}
	\nonumber \\
& = &
	- d
	( x_0 
	+ \frac{1}{2} x (\sigma + \Delta) x
	+ \frac{1}{2} \Delta x
	+ \varepsilon x)
	+ x (\omega + \sigma + \Delta) dx
	+ ( \alpha+\frac{1}{2} \Delta + \varepsilon ) dx
	\nonumber \\
& = &
	- d x_0 
	- x (\sigma + \Delta) dx
	- \frac{1}{2} \Delta dx
	- \varepsilon dx
	+ x (\omega + \sigma + \Delta) dx
	+ ( \alpha+\frac{1}{2} \Delta + \varepsilon ) dx
	\nonumber \\
& = &
	- d x_0 
	+ x \omega dx
	+ \alpha dx
	\nonumber \\
& = &
	A_{\omega, \alpha}
\end{eqnarray}
via pullback.
Thus the connections are classified by the equivalence relation
\begin{equation}
	( \omega, \alpha ) 
	\sim
	( \omega + \sigma + \Delta, \alpha + \frac{1}{2} \Delta + \varepsilon ),
	\qquad
	\sigma \in \mbox{EvenSym} (n,\Z);
	\:
	\Delta, \varepsilon \in \Z^n
	\label{classification}
\end{equation}
among 
$ ( \omega, \alpha ) \in \mbox{Mat} (n,\Z) \times \R^n $.

Next we define a covariant derivative 
of the equivariant function $ f $ by
\begin{equation}
	Df := df - 2 \pi i A f.
	\label{covariant derivative}
\end{equation}
Of course, on the right-hand side, $ i = \sqrt{-1} $.
The curvature form $ F $ is defined by
\begin{equation}
	F 
	:= dA 
	= 
	\sum_{j,k=1}^n \omega_{jk} dx_j \wedge dx_k
	= 
	\sum_{j,k=1}^n 
	\frac{1}{2} ( \omega_{jk} - \omega_{kj} ) dx_j \wedge dx_k,
	\label{curvature}
\end{equation}
which gives a constant magnetic field.
Hence the first Chern class is uniquely specified by
the integral antisymmetrized matrix $ ( \omega - {}^t \! \omega ) $.
It is known \cite{Hirzebruch}
that
an $ S^1 $-fiber bundle has a one-to-one correspondence with
the first Chern class.
Therefore, 
by choosing $ \omega \in \mbox{Mat} (n, \Z) $ appropriately,
we can construct any principal fiber bundles over $ T^n $
with the fiber $ S^1 $.

\section{Magnetic translation group}

Now we shall examine translation symmetry of 
the vector potential $ A $ of the uniform magnetic field.
In this section we shall give a precise definition of the MTG in $ T^n $
and express the MTG in a more concrete form.
We will prove that the MTG is
\begin{equation}
	S_A 
	= 
	(\R \times_\omega {\mit\Omega}^n) / (\Z \times_\omega \Z^n ),
	\label{S_A pre}
\end{equation}
where $ {\mit\Omega}^n $ is a subgroup of $ \R^n $ defined by
$
	{\mit\Omega}^n =
	\{ v \in \R^n \, | \,
	( \omega - {}^t \! \omega ) v \in \Z^n \}
$
and the group operation is taken in the sense of (\ref{twist product}).
This is one of the main results of this paper.

We begin by defining the MTG.
A vector $ v \in \R^n $ generates a translation of $ T^n $ by
\begin{equation}
	\tau_v : T^n \to T^n, \quad x \mapsto x+v.
	\label{translation of torus}
\end{equation}
When a map 
$ \widetilde{\tau}_v : P^{n+1}_\omega \to P^{n+1}_\omega $
satisfies the commutative diagram
\begin{eqnarray}
  \begin{array}{cccccc}
    & &          & S^1 &          & \\
    & & \swarrow &     & \searrow & \\
    & P^{n+1}_\omega && \mapright{\widetilde\tau}{} && P^{n+1}_\omega \\
    & \mapdown{\pi_\omega}{} & & & & \mapdown{}{\pi_\omega} \\
    & T^n && \mapright{\tau}{} && T^n 
   \end{array}
   \label{translation diagram}
\end{eqnarray}
the map $ \widetilde{\tau}_v $ is called a lift of the translation $ \tau_v $.
The lifted translations that leave the connection $ A $ invariant
form a group
\begin{equation}
	S_A
	:= 
	\{ \, \widetilde{\tau}_v : P^{n+1}_\omega \to P^{n+1}_\omega
	\: | \:
	v \in \R^n, \,
	\pi_\omega \circ \widetilde{\tau}_v = \tau_v \circ \pi_\omega;
	\:
	\widetilde{\tau}_v^* A = A \, \}.
	\label{MTG}
\end{equation}
We call it 
the stability group of $ A $, or
the magnetic translation group (MTG).

Let us write down the lifted translation in a more explicit form.
We use $ (x_0, x) \in \R^{n+1} $ as a coordinate of $ P^{n+1}_\omega $.
Since 
$ \pi_\omega \circ \widetilde{\tau}_v = \tau_v \circ \pi_\omega $,
the lift $ \widetilde{\tau}_v $ of (\ref{translation of torus}) must have 
the form
\begin{equation}
	\widetilde{\tau}_v : 
	(x_0, x) \mapsto 
	(x_0 + \theta(x_0,x,v), x+v).
	\label{lift of translation}
\end{equation}
To make $ \widetilde{\tau}_v $ commutative with the action of 
$ e^{2 \pi i w_0} \in S^1 $
the function $ \theta $ must satisfy
\begin{equation}
	x_0 + w_0 + \theta(x_0 + w_0,x,v) 
	=
	x_0 + \theta(x_0,x,v) + w_0,
\end{equation}
namely, $ \theta $ must satisfy
\begin{equation}
	\theta (x_0 + w_0,x,v) 
	=
	\theta (x_0,x,v)
	\label{S^1 commute}
\end{equation}
for any $ w_0 \in \R $.
Therefore,
the function $ \theta $ is independent of $ x_0 $.
To become a map of $ P^{n+1}_\omega $,
the map $ \widetilde{\tau}_v $ must send an orbit of the left-action of
$ \Z \times_\omega \Z^n $
to an orbit of the same group.
In other words,
for any $ (m_0, m) \in \Z \times_\omega \Z^n $
there must exist an element $ (m_0', m') \in \Z \times_\omega \Z^n $
that satisfies
\begin{equation}
	\widetilde{\tau}_v ( (m_0,m) \cdot (x_0,x))
	=
	(m_0',m') \cdot \widetilde{\tau}_v (x_0,x ).
\end{equation}
The above equation is rewritten as
$$
	( m_0 + x_0 + m \omega x + \theta(m+x,v), m+x+v)
	= 
	( m_0' + x_0 + \theta(x,v) + m' \omega (x+v), m'+x+v ),
$$
which is equivalent to a set of equations
\begin{eqnarray}
	m & = & m',
	\label{m=m'} \\
	m_0  + m \omega x + \theta(m+x,v)
	& = &
	m_0' + \theta(x,v) + m' \omega (x+v).
\end{eqnarray}
The last equation implies that
\begin{equation}
	\theta(m+x,v) - \theta(x,v) - m \omega v 
	= m_0' - m_0
	\in \Z
	\label{t condition}
\end{equation}
for any $ m \in \Z^n $.
In reverse order,
any function $ \theta(x,v) $ satisfying the condition
(\ref{t condition})
defines a lifted translation $ \widetilde{\tau}_v $ 
by (\ref{lift of translation}).
The lifted translation $ \widetilde{\tau}_v $ is actually
a combination of a spatial shift by $ v $
with a gauge transformation by $ \theta $.
Hence, we finish characterizing the lifted translations.

Let the lifted translation $ \widetilde{\tau}_v $
act on the connection form $ A $ of (\ref{connection})
via pull-back. Then it gives
\begin{eqnarray}
	\widetilde{\tau}_v^* A 
	& = &
	- ( dx_0 + d \theta )
	+ (x+v) \omega d(x+v) + \alpha d(x+v)
	\nonumber \\
	& = &
	A - d \theta + v \omega dx.
\end{eqnarray}
Hence, to leave the connection invariant as $ \widetilde{\tau}_v^* A = A $,
the function $ \theta $ must satisfy a differential equation
$ d \theta = v \omega dx $.
Thus we have
\begin{equation}
	\theta(x,v) = v \omega x + v_0
	\label{sol t}
\end{equation}
with a constant $ v_0 \in \R $.
To make $ \widetilde{\tau}_v $ a map of $ P^{n+1}_\omega $,
the function $ \theta $ must satisfy the condition (\ref{t condition}),
which requires that
\begin{equation}
	\theta(m+x,v) - \theta(x,v) - m \omega v 
	=  v \omega m - m \omega v 
	= -m ( \omega - {}^t \! \omega ) v
	\in \Z
	\label{sol t special}
\end{equation}
for any $ m \in \Z^n $.
Therefore, the vector $ v \in \R^n $ is required to satisfy
\begin{equation}
	( \omega - {}^t \! \omega ) v \in \Z^n.
	\label{v condition}
\end{equation}
We call the vector $ v $ satisfying (\ref{v condition}) 
a magnetic shift.
A set of the magnetic shifts is denoted by
\begin{equation}
	{\mit\Omega}^n :=
	\{ v \in \R^n \, | \,
	( \omega - {}^t \! \omega ) v \in \Z^n \}.
	\label{O}
\end{equation}
The set
$ {\mit\Omega}^n $ becomes an additive subgroup of $ \R^n $.
When the antisymmetrized matrix 
$ ( \omega - {}^t \! \omega ) $ is nondegenerated,
$ {\mit\Omega}^n $ is discrete.
The lifted translation $ \widetilde{\tau}_v $ 
defined by (\ref{lift of translation}) with (\ref{sol t})
becomes
\begin{equation}
	\widetilde{\tau}_v : 
	(x_0, x) \mapsto 
	(x_0 + \theta(x_0,x,v), x+v) =
	(x_0 + v \omega x + v_0, x+v) =
	(v_0, v ) \cdot (x_0, x),	
	\label{lift of translation'}
\end{equation}
and therefore the action of $ \widetilde{\tau}_v $ 
is identified with the action of 
$ (v_0, v) \in \R \times_\omega {\mit\Omega}^n $
on $ P^{n+1}_\omega $ from the left.
However, the subgroup
$ \Z \times_\omega \Z^n \subset \R \times_\omega {\mit\Omega}^n $
acts on $ P^{n+1}_\omega $ trivially.
Thus the stability group $ S_A $ of the connection $ A $ is identified as
\begin{equation}
	S_A 
	= 
	(\R \times_\omega {\mit\Omega}^n) / (\Z \times_\omega \Z^n ).
	\label{S_A}
\end{equation}
This is one of the main results of this paper.
Note that $ S_A $ is a central extension
of a compact Abelian group $ {\mit\Omega}^n / \Z^n $ by $ S^1 = \R / \Z $.

Actually, 
there is another way to characterize the group $ S_A $.
The group $ \R \times_\omega {\mit\Omega}^n $ is a normalizer of
$ N = \Z \times_\omega \Z^n $ in
$ G = \R \times_\omega \R^n $.
In other words, the subgroup $ H $ defined by
\begin{equation}
	H := \{ h \in G \, | \, \forall n \in N, h n h^{-1} \in N \}
	\label{nomalizer}
\end{equation}
coincides with $ \R \times_\omega {\mit\Omega}^n $.
The above statement is easily proved as follows.
A straightforward calculation yields
\begin{equation}
	( x_0, x ) \cdot ( m_0, m ) \cdot ( x_0, x )^{-1} 
	= 
	( m_0 + x ( \omega - {}^t \! \omega ) m, m ).
	 \label{normality}
\end{equation}
Therefore the necessary and sufficient condition for 
$ (x_0, x) \in \R \times_\omega \R^n $ 
to bring the above element into $ N = \Z \times_\omega \Z^n $ is 
that $ ( \omega - {}^t \! \omega ) x \in \Z^n $, 
or that $ x \in {\mit\Omega}^n $.
Thus 
$ N = \Z \times_\omega \Z^n $ is a normal subgroup of
$ H = \R \times_\omega {\mit\Omega}^n $,
and hence the quotient group $ S_A = H/N $ is well defined.

\section{Representations of the MTG in a three-torus}

A unitary representation theory of the MTG is significant
for spectral analyses of 
the Laplace operator and the Dirac operator
in a background gauge field.
In this section we examine a three-dimensional torus
and construct a complete set of representations.
This is another main result of this paper.
In the next section we will discuss an outline of
the representation theory of the MTG for arbitrary dimensions.

\subsection{Method}
The MTG was identified as
$
	S_A =
	(\R \times_\omega {\mit\Omega}^n) / (\Z \times_\omega \Z^n )
$
at (\ref{S_A}).
We would like to express the MTG in terms of generators and relations.
Here we concentrate on the three-dimensional torus.
Let us take the matrix
\begin{equation}
	\omega =
	\left(
	\begin{array}{ccc}
		0  & b_3 & -b_2 \\
		0  &  0  &  b_1 \\
		0  &  0  &   0    
	\end{array}
	\right)
	\label{omega3}
\end{equation}
with positive integers $ b_1 $, $ b_2 $, and $ b_3 $.
Then antisymmetrization of $ \omega $ yields
\begin{equation}
	\omega - {}^t \!\omega =
	\left(
	\begin{array}{rrr}
		 0   & b_3  & -b_2  \\
		-b_3 &  0   &  b_1 \\
		 b_2 & -b_1 &   0    
	\end{array}
	\right).
	\label{anti omega3}
\end{equation}
The characteristic equation of $ ( \omega - {}^t \!\omega ) $ is
\begin{equation}
	\det( \lambda - ( \omega - {}^t \!\omega ) )
	= 
	\lambda
	( \lambda^2 + b_1^2 + b_2^2 + b_3^2 ).
	\label{chara omega3}
\end{equation}
Hence, its eigenvalues are
\begin{equation}
	\lambda = 0, \, \pm i B
	\label{eigen B}
\end{equation}
with $ B := \sqrt{b_1^2 + b_2^2 + b_3^2} $.
We assume that $ B \ne 0 $.
The eigenspace for $ \lambda = 0 $ is spanned by
\begin{equation}
	b =
	\left(
	\begin{array}{c}
		b_1 \\
		b_2 \\
		b_3
	\end{array}
	\right).
	\label{0-vector}
\end{equation}
The action of 
$ ( \omega - {}^t \!\omega ) $ on a vector $ v \in \R^3 $
is equivalent to the vector product
$ ( \omega - {}^t \!\omega ) v = v \times b $.
The magnetic shift group (\ref{O}) now becomes
\begin{equation}
	{\mit\Omega}^3 =
	\{ v \in \R^3 \, | \,
	( \omega - {}^t \! \omega ) v \in \Z^3 \}.
	\label{O3}
\end{equation}
The linear subspace $ \R b $
spanned by $ b $ of (\ref{0-vector}) is a subgroup of $ {\mit\Omega}^3 $.
Let us define a generator $ e_0 $ by
\begin{eqnarray}
	D_0 & := & \GCD \{ b_1, b_2, b_3 \},
	\label{D0} \\
	e_0 & := & \frac{1}{D_0}
	\left(
		\begin{array}{r} b_1 \\ b_2 \\ b_3 \end{array}
	\right).
	\label{3generators}
\end{eqnarray}
Here the GCD is an abbreviation of
the greatest common divisor
while the LCM is an abbreviation of
the least common multiple.
It is obvious that
$ e_0 $ is in $ \Z^3 $ and that $ ( \omega - {}^t \! \omega ) e_0 = 0 $.
The vector $ e_0 $ is a minimal integral vector
in the sense that there is no real number $ s $
such that $ 0 < s < 1 $ and $ s e_0 \in \Z^3 $.
There exist other vectors $ e_1, e_2 \in \Q^3 $ that generate 
$ {\mit\Omega}^3 $ as
\begin{equation}
	{\mit\Omega}^3 =
	\R e_0 \oplus
	\Z e_1 \oplus
	\Z e_2.
	\label{dec omega3}
\end{equation}
Here $ \Q $ is the whole set of rational numbers.
Since (\ref{O3}) these generators $ e_1 $ and $ e_2 $ must satisfy
\begin{equation}
	( \omega - {}^t \! \omega ) e_1 , \,
	( \omega - {}^t \! \omega ) e_2 
	\in \Z^3.
	\label{Omega condition}
\end{equation}
These vectors $ \{ e_1, e_2 \} $ are minimal magnetic shifts
in the sense that there is no real number $ s $ such that
\begin{equation}
	0 < s < 1, \qquad
	s ( \omega - {}^t \! \omega ) e_i \in \Z^3
	\label{minimal}
\end{equation}
for each $ i = 1, 2 $.
Moreover, 
because $ e_1, e_2 \in \Q^3 $,
there are positive integers, $ \nu_1 $ and $ \nu_2 $, such that
\begin{equation}
	\nu_1 e_1 , \,
	\nu_2 e_2   \in \Z^3.
	\label{integ}
\end{equation}
We demand that 
the integers $ \{ \nu_1, \nu_2 \} $ are the smallest cycles
in the sense that there is no integer $ m $ such that
\begin{equation}
	0 < m < \nu_i, \qquad m e_i \in \Z^3
	\label{minimal int}
\end{equation}
for each $ i = 1, 2 $.
Consequently, the decomposition (\ref{dec omega3}) yields
\begin{equation}
	{\mit \Omega}^3 / \Z^3
	= (\R / \Z ) \oplus (\Z / \nu_1 \Z) \oplus (\Z / \nu_2 \Z).
	\label{decompo}
\end{equation}
Thus an arbitrary element $ g $ of 
$ S_A = (\R \times_\omega {\mit\Omega}^3) / (\Z \times_\omega \Z^3) $
is parametrized as
\begin{equation}
	g = (s, t e_0 + n_1 e_1 + n_2 e_2 ),
	\qquad
	s, t \in \R / \Z; \, 
	n_1  \in \Z / \nu_1 \Z; \,
	n_2  \in \Z / \nu_2 \Z.
\end{equation}
Let us examine the commutator (\ref{commutativity}).
It is clear that the element $ (s,0) $ commutes with any element.
Since $ ( \omega - {}^t \! \omega ) e_0 = 0 $,
the element $ (0,t e_0) $ also commutes with any element.
On the other hand, $ (0, e_1) $ and $ (0, e_2) $ produce 
a nonvanishing commutator
\begin{equation}
	(0,e_1) \cdot (0,e_2) \cdot (0,e_1)^{-1} \cdot (0,e_2)^{-1} 
	= ( \gamma, 0 )
	\label{beta commutator}
\end{equation}
with 
\begin{eqnarray}
	\gamma 
	:= 
	e_1 ( \omega - {}^t \! \omega ) e_2
	= 
	e_1 \cdot ( e_2 \times b ).
	\label{beta}
\end{eqnarray}
{}From (\ref{Omega condition}) and (\ref{integ}) we can see that
\begin{equation}
	\nu_1 \gamma, \, \nu_2 \gamma \in \Z.
	\label{nu gamma}
\end{equation}
Hence $ \gamma $ is a rational number.
Let $ d $ be the greatest common divisor of $ \nu_1 $ and $ \nu_2 $.
If we put
$ \nu_1 = d p_1 $ and $ \nu_2 = d p_2 $, then
$ p_1 $ and $ p_2 $ are mutually prime.
The above equation (\ref{nu gamma}) implies
that $ d \gamma $ is an integer.
So we have
\begin{equation}
	d := \GCD \{ \nu_1, \nu_2 \},
	\qquad
	\ell := d \gamma \in \Z.
	\label{rationality}
\end{equation}

Before constructing the representation of the MTG,
we need to know how the generators generate
an arbitrary element of the MTG.
{}From the multiplication rule of the group $ \R \times_\omega \R^n $
we deduce that for $ x, y \in \R^n $
\begin{eqnarray}
	\left( \frac{1}{2} x \omega x, x \right) \cdot 
	\left( \frac{1}{2} y \omega y, y \right) 
& = &
	\left( 
	\frac{1}{2} x \omega x + \frac{1}{2} y \omega y + x \omega y, x+y 
	\right) 
	\nonumber \\
& = &
	\left( 
	\frac{1}{2} x \omega y - \frac{1}{2} y \omega x
	+ \frac{1}{2} (x+y) \omega (x+y), x+y 
	\right) 
	\nonumber \\
& = &
	\left( \frac{1}{2} x ( \omega - {}^t \! \omega ) y, 0 \right) \cdot
	\left( \frac{1}{2} (x+y) \omega (x+y), x+y \right)
	\;
	\label{additivity}
\end{eqnarray}
and
\begin{equation}
	\left( \frac{1}{2} x \omega x,  x \right)^{-1} =
	\left( \frac{1}{2} x \omega x, -x \right).
	\label{inverse xx}
\end{equation}
Iteration of (\ref{additivity}) yields
\begin{equation}
	\left( \frac{1}{2} x \omega x,  x \right)^n =
	\left( \frac{1}{2} n^2 x \omega x, nx \right),
	\qquad n \in \Z.
	\label{n}
\end{equation}
Furthermore, (\ref{additivity}) implies
\begin{equation}
	\left( \frac{1}{2} s^2 x \omega x,  s x \right) \cdot
	\left( \frac{1}{2} t^2 x \omega x,  t x \right) 
	=
	\left( \frac{1}{2} (s+t)^2 x \omega x,  (s+t) x \right),
	\qquad s, t \in \R.
	\label{real}
\end{equation}
By a tedious calculation we can show
\begin{equation}
	(s, x+y+z) 
	=
	\left( s - X, 0 \right) \cdot
	\left( \frac{1}{2} x \omega x,  x \right) \cdot
	\left( \frac{1}{2} y \omega y,  y \right) \cdot
	\left( \frac{1}{2} z \omega z,  z \right)
	\label{3 inverse}
\end{equation}
with
\begin{equation}
	X =
	  \frac{1}{2} (x+y+z) \omega (x+y+z)
	+ \frac{1}{2} x (\omega - {}^t \! \omega) y
	+ \frac{1}{2} x (\omega - {}^t \! \omega) z
	+ \frac{1}{2} y (\omega - {}^t \! \omega) z. \qquad
	\label{X}
\end{equation}
Thus an arbitrary element of the MTG is expressed as
\begin{eqnarray}
	g 
& = &
	(s, t e_0 + n_1 e_1 + n_2 e_2) 
	\nonumber \\
& = &
	\left( 
		s 
		- \frac{1}{2} 
		(t e_0 + n_1 e_1 + n_2 e_2) \omega 
		(t e_0 + n_1 e_1 + n_2 e_2) 
		- \frac{1}{2} \gamma n_1 n_2,
		0 
	\right) 
	\nonumber \\ &&
	\cdot
	\left( \frac{1}{2} t^2 e_0 \omega e_0, t e_0 \right) \cdot
	\left( \frac{1}{2} e_1 \omega e_1, e_1 \right)^{n_1} \cdot
	\left( \frac{1}{2} e_2 \omega e_2, e_2 \right)^{n_2} 
	\nonumber \\
& = &
	\phi (s-X) \cdot g_0(t) \cdot (g_1)^{n_1} \cdot (g_2)^{n_2},
	\label{4dec}
\end{eqnarray}
which is a product of the generators
\begin{eqnarray}
	\phi(s)& := & (s,0), \\
	g_0(t) & := & \left( \frac{1}{2} t^2 e_0 \omega e_0, t e_0 \right), \\
	g_1    & := & \left( \frac{1}{2} e_1 \omega e_1, e_1 \right), \\
	g_2    & := & \left( \frac{1}{2} e_2 \omega e_2, e_2 \right).
\end{eqnarray}
These generators satisfy the relations
\begin{eqnarray}
&&	\phi(s) \cdot \phi(t) = \phi(s+t), \label{phi}
	\label{phi(s)} \\
&&	\phi(1) = 1, 
	\label{phi(1)} \\
&&	g_0(s) \cdot g_0(t) = g_0(s+t), \label{g} \\
&&	g_0 (1) = \phi \left( \frac{1}{2} z_0 \right), \\
&&	(g_1)^{\nu_1} = \phi \left( \frac{1}{2} z_1 \right), \\
&&	(g_2)^{\nu_2} = \phi \left( \frac{1}{2} z_2 \right), \\
&&	g_1 \cdot g_2 \cdot g_1^{-1} \cdot g_2^{-1} = \phi(\gamma) 
	\label{g comm}
\end{eqnarray}
and other trivial commutators.
Here we have defined $ \{z_0, z_1, z_2 \} $ by
\begin{eqnarray}
	z_0 
& := &
	e_0 \omega e_0
	=
	\left( \frac{1}{D_0} \right)^2 b_1 b_2 b_3,
	\label{z0}
	\\
	z_1 
& := &
	\nu_1^2 e_1 \omega e_1,
	\qquad
	z_2 
 := 
	\nu_2^2 e_2 \omega e_2.
	\label{z2}
\end{eqnarray}
Because $ e_0 $ is an integral vector and $ \omega $ is an integral matrix,
$ z_0 $ is an integer.
Furthermore, (\ref{integ}) implies that $ z_1 $ and $ z_2 $ are also integers.
In reverse order,
the generators 
$ \{ \phi(s), g_0(t), g_1, g_2 \} $
and their relations (\ref{phi})-(\ref{g comm}) determine the MTG uniquely.
These generators with the relations form the MTG 
in a constructive manner.
Consequently, the MTG in $ T^3 $ is completely characterized by
the set of parameters 
$ ( z_0, z_1, z_2, \nu_1, \nu_2, \gamma ) $,
where 
$ \{ z_0, z_1, z_2, \nu_1, \nu_2 \} $ are integers and
$ \gamma $ is a rational number constrained 
by the condition (\ref{nu gamma}).

Now we discuss the representation theory of the MTG exhaustively.
The space of functions $ \{ f : \R^{n+1} \to \C \} $ provides
the regular representation of
the group $ \R \times_\omega \R^ n $ via
\begin{eqnarray}
	U(v_0, v) f(x_0, x) 
	& := & f( (v_0,v)^{-1} \cdot (x_0,x) )
	\nonumber \\
	& = & f( (-v_0 + v \omega v, -v) \cdot (x_0,x) )
	\nonumber \\
	& = & f( x_0 - v_0 + v \omega v -v \omega x, x-v ).
	\label{left representation}
\end{eqnarray}
We restrict the representation $ U $ on the space of equivariant functions,
which are constrained by (\ref{invariance}) and (\ref{equivariance}).
Then we have
\begin{equation}
	U(v_0, v) f(x_0, x) 
	= 
	e^{2 \pi i ( v_0 - v \omega v + v \omega x ) }
	f( x_0 , x-v ),
	\label{left rep on equi func}
\end{equation}
which reproduces the unitary transformations (\ref{U_x}) and (\ref{U_y})
when the twisting matrix (\ref{T^2}) is taken.
Particularly $ (v_0, 0) $ is represented by
\begin{equation}
	U(v_0, 0) f(x_0, x) 
	= e^{ 2 \pi i v_0} f( x_0, x ).
	\label{left representation of S1}
\end{equation}
Hence the representation $ U $ induces an isomorphism of 
(\ref{phi(s)}) and (\ref{phi(1)}) by
\begin{equation}
	U(\phi(s)) = e^{ 2 \pi i s}.
	\label{Uphi}
\end{equation}
Moreover, if we put
\begin{equation}
	U_0 (t) :=  U ( g_0(t) ), \quad
	U_1     := U ( g_1 ), \quad
	U_2     := U ( g_2 ),
	\label{UU}
\end{equation}
they satisfy
\begin{eqnarray}
&&	U_0(s) \, U_0(t) = U_0(s+t), \label{U0} \\
&&	U_0 (1) =       e^{ \pi i z_0 }, \\
&&	(U_1)^{\nu_1} = e^{ \pi i z_1 }, \label{U1n1} \\
&&	(U_2)^{\nu_2} = e^{ \pi i z_2 }, \label{U2n2} \\
&&	U_1 \, U_2 \, U_1^{-1} \, U_2^{-1} = e^{2 \pi i \gamma} 
	\label{U comm}
\end{eqnarray}
since $ U $ is a homomorphism of the relations (\ref{g})-(\ref{g comm}).

An irreducible representation of $ U_0(t) $ is 
labeled by an integer $ q_0 $ and defined by
\begin{equation}
	U_0 (t) | q_0 \ket
	= e^{2 \pi i ( q_0 + \frac{1}{2} z_0 )t } | q_0 \ket.
	\label{rep U0}
\end{equation}
On the other hand, 
to construct a representation of the algebra generated by $ U_1 $ and $ U_2 $,
we introduce a set of orthogonal vectors
$ \{ | q_1, q_2 \ket \, | \,
 q_1 \in \Z / \nu_1 \Z, \,
 q_2 \in \Z / \nu_2 \Z \} $.
We assume identification
$ | q_1, q_2 \ket = | q_1 + k_1 \nu_1, q_2 + k_2 \nu_2 \ket $ 
for any $ k_1, k_2 \in \Z $.
Let the operators $ U_1 $ and $ U_2 $ act on them by
\begin{eqnarray}
	&&
	U_1 | q_1, q_2 \ket 
	= e^{2 \pi i (q_1 + \frac{1}{2} z_1) / \nu_1} 
	| q_1, q_2 \ket,
	\label{rep U1} \\
	&&
	U_2 | q_1, q_2 \ket 
	= e^{2 \pi i (q_2 + \frac{1}{2} z_2) / \nu_2} 
	| q_1 + \nu_1 \gamma, q_2 \ket.
	\label{rep U2}
\end{eqnarray}
The step of $ q_1 $ generated by $ U_2 $ is
\begin{equation}
	\Delta q_1 
	:=\nu_1 \gamma 
	= d p_1 \, \frac{\ell}{d} 
	= p_1 \ell.
	\label{q_1 step}
\end{equation}
Then the fundamental relations,
(\ref{U1n1}), (\ref{U2n2}), and (\ref{U comm}),
are satisfied as
\begin{eqnarray}
	(U_1)^{\nu_1} | q_1, q_2 \ket 
& = &
	e^{2 \pi i (q_1 + \frac{1}{2} z_1)} 
	| q_1, q_2 \ket
	\nonumber \\
& = &
	e^{2 \pi i \frac{1}{2} z_1 } 
	| q_1, q_2 \ket,
	\label{rel U1} 
\\
	(U_2)^{\nu_2} | q_1, q_2 \ket 
& = &
	e^{2 \pi i (q_2 + \frac{1}{2} z_2) } 
	| q_1 + \nu_1 \nu_2 \gamma, q_2 \ket
	\nonumber \\
& = &
	e^{2 \pi i \frac{1}{2} z_2 } 
	| q_1, q_2 \ket
	\qquad (\mbox{because}\; \nu_2 \gamma \; \mbox{is an integer}),
	\label{rel U2} 
\\
	U_1 \, U_2 \, U_1^{-1} \, U_2^{-1} | q_1, q_2 \ket
& = &	
	U_1 \, U_2 \, U_1^{-1} \,
	e^{-2 \pi i (q_2 + \frac{1}{2} z_2) / \nu_2}
	| q_1 - \nu_1 \gamma, q_2 \ket
	\nonumber \\
& = &	
	U_1 \, U_2 \,
	e^{-2 \pi i(q_1 - \nu_1 \gamma +\frac{1}{2} z_1) / \nu_1} 
	e^{-2 \pi i (q_2 + \frac{1}{2} z_2) / \nu_2}
	| q_1 - \nu_1 \gamma, q_2 \ket
	\nonumber \\
& = &	
	U_1 \,
	e^{-2 \pi i(q_1 - \nu_1 \gamma +\frac{1}{2} z_1) / \nu_1} 
	| q_1, q_2 \ket
	\nonumber \\
& = &	
	e^{ 2 \pi i \gamma} 
	| q_1, q_2 \ket.
	\label{rel com}
\end{eqnarray}
Thus the basis
$ \{ | q_1, q_2 \ket \} $
spans a representation space of the algebra generated by $ U_1 $ and $ U_2 $.
This representation space is reducible generally.
We can see that the action of $ U_2 $ is cyclic. 
Namely, if we put
\begin{equation}
	c 
	:= \frac{ \LCM \{\Delta q_1, \nu_1 \} }{ \Delta q_1 }
	 = \frac{ \LCM \{ p_1 \ell, p_1 d \} }{ p_1 \ell }
	 = \frac{ \LCM \{ \ell, d \} }{ \ell }
	 = \frac{d}{ \GCD \{ \ell, d \} },
	\label{cycle}
\end{equation}
then
\begin{equation}
	c \, \Delta q_1
	= \LCM \{ \Delta q_1, \nu_1 \} 
	\label{cyclic}
\end{equation}
is an integral multiple of $ \nu_1 $ and therefore
the $ U_2 $ action (\ref{rep U2}) iterated by $ c $ times gives
\begin{equation}
	(U_2)^c | q_1, q_2 \ket 
	= 
	e^{2 \pi i (q_2 + \frac{1}{2} z_2) c / \nu_2} 
	| q_1 + c \, \Delta q_1, q_2 \ket
	= 
	e^{2 \pi i (q_2 + \frac{1}{2} z_2) c / \nu_2} 
	| q_1, q_2 \ket.
\end{equation}
Moreover,
\begin{eqnarray}
	&&
	\frac{\nu_1}{c} 
	 = \frac{ \nu_1 \GCD \{ \ell, d \} }{d}
	 = \frac{ d p_1 \GCD \{ \ell, d \} }{d}
	 = p_1 \GCD \{ \ell, d \},
	\label{mod1}
	\\ &&
	\frac{\nu_2}{c} 
	 = \frac{ \nu_2 \GCD \{ \ell, d \} }{d}
	 = \frac{ d p_2 \GCD \{ \ell, d \} }{d}
	 = p_2 \GCD \{ \ell, d \}
	\label{mod2} 
\end{eqnarray}
are integers.
Therefore,
each choice of 
$ q_1 \in \Z $ modulo $ (\nu_1 / c) \Z $ and
$ q_2 \in \Z $ modulo $ (\nu_2 / c) \Z $
specifies one of inequivalent irreducible representations.
Consequently,
the dimension of the irreducible representation is 
\begin{equation}
	\mbox{dimension} 
	= c
	= \frac{ d }{ \GCD \{ d, \ell \}}.
	\label{dim}
\end{equation}
On the other hand, 
the number of inequivalent representations for a fixed $ q_0 $ is
\begin{equation}
	\# \mbox{ineq. irr. rep.} 
	= \frac{ \nu_1 }{c} \cdot \frac{ \nu_2 }{c}
	= p_1 p_2 ( \GCD \{ d, \ell \} )^2.
	\label{num}
\end{equation}
These numbers give a number
\begin{equation}
	(\mbox{dimension})^2 \times 
	(\# \mbox{ineq. irr. rep.})
	= c^2 \times \frac{ \nu_1 \nu_2 }{c^2}
	= \nu_1 \nu_2,
	\label{complete}
\end{equation}
which coincides with the dimension of
the algebra generated by $ U_1 $ and $ U_2 $,
as required by the Peter-Weyl theory on group representation \cite{Serre}.
Thus we have obtained the complete set 
of irreducible representations of the algebra.

In summary, an irreducible representation of the MTG 
in the three-dimensional torus is specified
by 
\begin{equation}
	\chi = (q_0, [q_1], [q_2] ) \in 
	\Z \times \Z_{(\nu_1/c)} \times \Z_{(\nu_2/c)}.
	\label{label}
\end{equation}
Using the decomposition (\ref{4dec})
we have
\begin{eqnarray}
	&&
	U(s, t e_0 + n_1 e_1 + n_2 e_2 ) 
	| q_0, q_1, q_2 \ket
	\nonumber \\
& = &
	e^{2 \pi i ( s - X ) }
	\,
	U_0 (t)
	\,
	(U_1)^{n_1}
	\,
	(U_2)^{n_2}
	| q_0, q_1, q_2 \ket
	\nonumber \\
& = &
	e^{2 \pi i 
	\{ 
		(s - X)
		+ ( q_0 + \frac{1}{2} z_0 ) t 
		+ ( q_1+ \gamma \nu_1 n_2 + \frac{1}{2} z_1 ) n_1 / \nu_1
		+ ( q_2 + \frac{1}{2} z_2 ) n_2 / \nu_2
	\} }
	| q_0, q_1 + \gamma \nu_1 n_2, q_2 \ket
	\;
	\label{3 representation}
\end{eqnarray}
with $ X $ evaluated as
\begin{eqnarray}
	X
& = &
	\frac{1}{2} 
		(t e_0 + n_1 e_1 + n_2 e_2) \omega 
		(t e_0 + n_1 e_1 + n_2 e_2) 
	+ \frac{1}{2} \gamma n_1 n_2.
\end{eqnarray}

\subsection{Examples in the three-dimensional torus}
Here we apply the previous method of representation of the MTG
to three examples of magnetic fields in $ T^3 $.

The first example is a magnetic field parallel to the $ x_3 $-axis,
\begin{equation}
	b 
	=
	\left(
		\begin{array}{r} b_1 \\ b_2 \\ b_3 \end{array}
	\right)
	=
	\left(
		\begin{array}{r} 0 \\ 0 \\ \nu \end{array}
	\right)
	\label{ex1:b}
\end{equation}
with a positive integer $ \nu $. 
The generators (\ref{3generators}) of the MTG are 
\begin{equation}
	e_0 
	= 
	\left(
		\begin{array}{r} 0 \\ 0 \\ 1 \end{array}
	\right),
	\quad
	e_1 
	= 
	\frac{1}{\nu}
	\left(
		\begin{array}{r} 1 \\ 0 \\ 0 \end{array}
	\right),
	\quad
	e_2 
	= 
	\frac{1}{\nu}
	\left(
		\begin{array}{r} 0 \\ 1 \\ 0 \end{array}
	\right).
\end{equation}
The vectors $ \{ e_1, e_2 \} $ reproduce the discrete magnetic shifts
(\ref{discrete}) in the plane perpendicular to the magnetic field.
The cycles (\ref{integ}) of $ e_1 $ and $ e_2 $ are found to be
\begin{equation}
	\nu_1 = \nu,
	\qquad
	\nu_2 = \nu,
	\label{ex1:alpha2} 
\end{equation}
respectively. 
Using them we evaluate the parameters of the MTG as
\begin{eqnarray}
	&&
	d 
	= \GCD \{ \nu_1, \nu_2 \}
	= \GCD \{ \nu, \nu \}
	= \nu,
	\label{ex1:d}
\\	&&
	\gamma
	= e_1 ( \omega - {}^t \! \omega ) e_2
	= e_1 \cdot ( e_2 \times b )
	= \frac{1}{\nu},
	\label{ex1:beta}
\\	&&
	\ell
	= d \gamma
	= \nu \frac{1}{\nu}
	= 1,
	\label{ex1:f}
\\	&&
	z_0 = z_1 = z_2 = 0.
	\label{ex1:zs}
\end{eqnarray}
The size and the number of irreducible representations are
\begin{eqnarray}
	&&
	\mbox{dimension} 
	= c 
	= \frac{d}{ \GCD \{ \ell, d \} }
	= \frac{\nu}{ \GCD \{ 1, \nu \} }
	= \frac{\nu}{ 1 }
	= \nu,
	\label{ex1:c}
\\	&&
	\# \mbox{ineq. irr. rep.} 
	= \frac{ \nu_1 \nu_2 }{c^2}
	= \frac{ \nu^2 }{\nu^2}
	= 1.
	\label{ex1:num}
\end{eqnarray}
In this case 
(\ref{rep U1}) and (\ref{rep U2})
reproduce the representations
(\ref{rep Ux}) and (\ref{rep Uy}) in $ T^2 $.

The second example is a magnetic field 
perpendicular to the $ x_1 $-axis
and lying in the middle of the $ x_2 $-  and $ x_3 $-axes,
\begin{equation}
	b 
	=
	\left(
		\begin{array}{r} b_1 \\ b_2 \\ b_3 \end{array}
	\right)
	=
	\left(
		\begin{array}{r} 0 \\ \nu \\ \nu \end{array}
	\right)
	\label{ex2:b}
\end{equation}
with a positive integer $ \nu $. 
The generators of the MTG are chosen as
\begin{equation}
	e_0 
	= 
	\left(
		\begin{array}{r} 0 \\ 1 \\ 1 \end{array}
	\right),
	\quad
	e_1 
	= 
	\frac{1}{\nu}
	\left(
		\begin{array}{r} 0 \\ 0 \\ 1 \end{array}
	\right),
	\quad
	e_2 
	= 
	\frac{1}{\nu}
	\left(
		\begin{array}{r} 1 \\ 0 \\ 0 \end{array}
	\right).
	\label{ex2:e3}
\end{equation}
The cycles of $ e_1 $ and $ e_2 $ are 
\begin{equation}
	\nu_1 = \nu,
	\quad
	\nu_2 = \nu
	\label{ex2:alpha2} 
\end{equation}
and other parameters of the MTG are also evaluated as
\begin{eqnarray}
	&&
	d = \nu,
	\quad
	\gamma = \frac{1}{\nu},
	\quad
	\ell= 1,
	\quad
	z_0 = z_1 = z_2 = 0,
	\quad
	\label{ex2:zs}
\\	&&
	\mbox{dimension} 
	= c 
	= \nu,
	\label{ex2:c}
\\	&&
	\# \mbox{ineq. irr. rep.} 
	= \frac{ \nu_1 \nu_2 }{c^2}
	= 1.
	\label{ex2:num}
\end{eqnarray}

The third example is a magnetic field in the direction of $ (1, 1, 1) $,
\begin{equation}
	b 
	=
	\left(
		\begin{array}{r} b_1 \\ b_2 \\ b_3 \end{array}
	\right)
	=
	\left(
		\begin{array}{r} \nu \\ \nu \\ \nu \end{array}
	\right)
	\label{ex3:b}
\end{equation}
with a positive integer $ \nu $.
A calculation similar to the previous ones gives a series of parameters.
Here we show only the results
\begin{eqnarray}
	&& 
	e_0 
	= 
	\left(
		\begin{array}{r} 1 \\ 1 \\ 1 \end{array}
	\right),
	\quad
	e_1 
	= 
	\frac{1}{\nu}
	\left(
		\begin{array}{r} 1 \\ 0 \\ 0 \end{array}
	\right),
	\quad
	e_2 
	= 
	\frac{1}{\nu}
	\left(
		\begin{array}{r} 0 \\ 1 \\ 0 \end{array}
	\right),
	\label{ex3:e3}
\\	&&
	\nu_1 = \nu,
	\quad
	\nu_2 = \nu,
	\quad
	d = \nu,
	\quad
	\gamma = \frac{1}{\nu},
	\quad
	\ell = 1,
	\quad
	z_0 = \nu, 
	\quad
	z_1 = z_2 = 0,
	\label{ex3:zs}
\\	&&
	\mbox{dimension} = c = \nu
	\label{ex3:c}
\\	&&
	\# \mbox{ineq. irr. rep.} = \frac{ \nu_1 \nu_2 }{c^2} = 1.
	\label{ex3:num}
\end{eqnarray}

\section{Representation theory of the MTG in an $ n $-torus}

Here we describe how to characterize the MTGs in an $ n $-dimensional torus.
We can choose generators
$ \{ e_1, e_2, \cdots, e_l, f_1, f_2, \cdots, f_l, g_1, g_2, \cdots, g_m \} $
$ (2l+m=n) $ of the magnetic shift group $ {\mit\Omega}^n $ 
such that
\begin{eqnarray}
	&&
	e_i, f_i \in \Q^n,
	\quad
	( \omega - {}^t \! \omega ) e_i,
	( \omega - {}^t \! \omega ) f_i \in \Z^n,
	\quad
	e_i ( \omega - {}^t \! \omega ) f_j = \gamma_i \delta_{ij}
	\nonumber 
\\ && 
	\qquad \qquad \qquad \qquad \qquad \qquad \quad
	(i,j=1, \cdots, l)
	\label{gamma_i}
\\
	&&
	g_k \in \Z^n, 
	\quad
	( \omega - {}^t \! \omega ) g_k = 0
	\quad
	(k=1, \cdots, m)
\end{eqnarray}
with nonzero rational numbers $ \gamma_i \in \Q $.
The vectors
$ \{ g_1, \cdots, g_m \} $
are demanded to be minimal integral vectors in the sense that
there is no real number $ s $ satisfying
\begin{equation}
	0 < s < 1, \quad s g_k \in \Z^n
\end{equation}
for each $ k = 1, \cdots, m $.
Let $ \{ \mu_i, \nu_i \} \, (i=1,\cdots,l) $ 
be smallest positive integers such that
\begin{equation}
	\mu_i e_i, \nu_i f_i \in \Z^n
	\label{integral}
\end{equation}
and that there are no integers $ \{ m_i, n_i \} $ satisfying
\begin{eqnarray}
	&&
	0 < m_i < \mu_i, \quad m_i e_i \in \Z^n,
	\\
	&&
	0 < n_i < \nu_i, \quad n_i f_i \in \Z^n.
\end{eqnarray}
Equations (\ref{gamma_i}) and (\ref{integral}) imply that
$ \gamma_i \mu_i $ and $ \gamma_i \nu_i $ are integers.
Thus, by putting
\begin{equation}
	d_i := \GCD \{ \mu_i, \nu_i \},
\end{equation}
we can see that $ \gamma_i d_i $ is an integer.
Consequently, the group of translations (\ref{O}) is decomposed as
\begin{equation}
	{\mit\Omega}^n =
	\Z e_1 \oplus \cdots \oplus \Z e_l \oplus
	\Z f_1 \oplus \cdots \oplus \Z f_l \oplus
	\R g_1 \oplus \cdots \oplus \R g_m 
\end{equation}
and the MTG (\ref{S_A}) is expressed as
\begin{equation}
	S_A \cong
	S^1 \times_\omega
	( 
	\Z_{\mu_1} \times \cdots \times \Z_{\mu_l} \times
	\Z_{\nu_1} \times \cdots \times \Z_{\nu_l} \times
	T^m 
	).
	\label{MTG Tn}
\end{equation}

Finally, we describe an outline of 
the representation theory of the MTGs in an $ n $-dimensional torus.
Let us define integers $ x_i, y_i, z_j $ by
\begin{eqnarray}
	x_i & := & \mu_i^2 e_i \omega e_i, 
	\\
	y_i & := & \nu_i^2 f_i \omega f_i \qquad (i=1,\cdots,l),
	\\
	z_k & := & g_k \omega g_k \qquad (k=1,\cdots,m).
\end{eqnarray}
Generators of the MTG (\ref{MTG Tn}) are represented 
by a set of unitary operators
\begin{eqnarray}
	T(s) & := & U (s,0), \qquad s \in \R,
	\\
	U_i & := & U \left( \frac{1}{2} e_i \omega e_i, e_i \right),
	\\
	V_i & := & U \left( \frac{1}{2} f_i \omega f_i, f_i \right),
	\\
	W_k(t) & := & U \left( \frac{1}{2} t^2 g_k \omega g_k, t g_k \right),
	\qquad t \in \R.
\end{eqnarray}
They satisfy the equations
\begin{eqnarray}
	T(s) T(t) & = & T(s+t),
	\\
	T(1) & = & 1,
	\\
	(U_i)^{\mu_i} & = & T ( x_i/2 ),
	\\
	(V_i)^{\nu_i} & = & T ( y_i/2 ),
	\\
	U_i \, V_i \, U_i^{-1} \, V_i^{-1} & = & T(\gamma_i),
	\\
	W_k(s) W_k(t) & = & W_k(s+t),
	\\
	W_k(1) & = & T ( z_k/2 ),
\end{eqnarray}
and other trivial commutators.
These equations for the $ n $-torus are generalization of
the equations (\ref{phi(s)})-(\ref{g comm}) for the three-torus.
A representation space is spanned by the basis vectors
\begin{equation}
	| \lambda, p, q, r \ket
	=
	| \lambda, 
	p_1, p_2, \cdots, p_l, 
	q_1, q_2, \cdots, q_l, 
	r_1, r_2, \cdots, r_m \ket
	\label{basis}
\end{equation}
labeled by
$ \lambda \in \Z $,
$ p_i \in \Z_{\mu_i} $,
$ q_i \in \Z_{\nu_i} $, and
$ r_k \in \Z $.
The generators act on the basis vectors according to
\begin{eqnarray}
	T(s) 
	| \lambda, p, q, r \ket
	& = & 
	e^{2 \pi i \lambda s}
	| \lambda, p, q, r \ket,
\\
	U_i
	| \lambda, p, q, r \ket
	& = &
	e^{ \pi i \lambda (2 p_i + x_i) /\mu_i}
	| \lambda, p, q, r \ket,
\\
	V_i
	| \lambda, p, q, r \ket
	& = &
	e^{ \pi i \lambda (2 q_i + y_i) /\nu_i}
	| \lambda, p_1, \cdots, p_i + \gamma_i \mu_i, \cdots, p_l, q , r \ket,
\\
	W_k(t) 
	| \lambda, p, q, r \ket
	& = &
	e^{ \pi i \lambda (2 r_k + z_k) t }
	| \lambda, p, q, r \ket.
\end{eqnarray}
These are generalization of
(\ref{Uphi}) and (\ref{rep U0})-(\ref{rep U2}).
The cycle of $ V_i $ is given by
\begin{equation}
	c_i
	:= \frac{ \LCM \{ \gamma_i \mu_i, \mu_i \} }{ \gamma_i \mu_i }
	 = \frac{ \LCM \{ \gamma_i d_i, d_i \} }{ \gamma_i d_i }
	 = \frac{ d_i }{ \GCD \{ \gamma_i d_i, d_i \} }.
	\label{cyclei}
\end{equation}
Hence an irreducible representation is labeled by
\begin{eqnarray}
	&&
	\chi = 
	( \lambda, [p_1], \cdots, [p_l], [q_1], \cdots, [q_l], r_1, \cdots, r_m )
	\nonumber \\ && \quad
	\in
	\Z \times 
	\Z_{(\mu_1/c_1)} \times \cdots \times
	\Z_{(\mu_l/c_l)} \times 
	\Z_{(\nu_1/c_1)} \times \cdots \times
	\Z_{(\nu_l/c_l)} \times 
	\Z^m.
	\label{label for T^n}
\end{eqnarray}
The dimension of the irreducible representation is 
\begin{equation}
	\mbox{dimension} =
	\prod_{i=1}^l c_i
	\label{dim for T^n}
\end{equation}
and the number of inequivalent representations is
\begin{equation}
	\# \mbox{ineq. irr. rep.} 
	= \prod_{i=1}^l \frac{ \mu_i \nu_i }{c_i^2}
\end{equation}
for fixed $ (\lambda,r_1,\cdots,r_m) \in \Z^{m+1} $.

\section{Conclusion}

Let us summarize our discussions.
We began this paper with a discussion on
symmetry of a charged particle in a uniform magnetic field.
We saw that
the quantum system in $ T^2 $ has 
a discrete noncommutative translation symmetry.
The symmetry is characterized by a central extension of a cyclic group.

In the following part of this paper
we introduced a noncommutative product into $ \R^{n+1} $.
Using the group structure,
we defined the magnetic fiber bundles $ P^{n+1}_\omega $,
which is a fiber bundle over $ T^n $ with a fiber $ S^1 $.
Then we showed that the set of magnetic fiber bundles is classified 
by the quotient space of integral matrices 
$ \mbox{Mat} (n,\Z) / \mbox{Sym} (n,\Z) $.
We introduced connections into the fiber bundles and classified them by
$ \mbox{Mat} (n,\Z) \times \R^n / \mbox{Sym} (n,\Z) \times \Z^n $
as showed in (\ref{classification}).
The lifted translations leaving the connection invariant
form the magnetic translation group of (\ref{MTG}).
We characterized the MTG 
by (\ref{S_A}) with (\ref{O}).
This characterization of the MTG is one of main results of this paper.
We found that
the magnetic shift group $ {\mit\Omega}^n $ is discrete
when the characteristic matrix $ (\omega - {}^t \! \omega) $ is nondegenerated.

In the rest of paper
we discussed the representation theory of the MTG for $ T^3 $ in detail
and applied it to a few examples.
The dimensions of an irreducible unitary representation is given by $ c $ in
(\ref{dim})
and each irreducible representation is labeled by $ \chi $ in
(\ref{label}).
These results may be useful for application to the electron system
in a lattice in an inclined magnetic field.
We briefly described the representation theory of the MTG for $ T^n $
and summarized the result in
(\ref{label for T^n}) and (\ref{dim for T^n}).

Here we would like to mention remaining problems.
It is desirable to apply the representation theory of the MTG
to spectral analyses of the Laplace and Dirac operators.
Originally the spectral problem of the quantum mechanics in a torus
motivated this study.
For this application
the Peter-Weyl theory on group representation will play an essential role.
In the next study
we would like to pursue the analysis of the Laplace operator 
in the torus with a magnetic field.
Moreover, an equilateral torus admits discrete transformations
that exchange vertices of the torus 
and that leave the metric and the magnetic field invariant.
It is also desirable to include such discrete transformations into the MTG
for the complete spectral analysis.

By developing the theory of the MTG
we will find its applications to physics.
Inclusion of the supersymmetry into the MTG 
is an interesting direction for the future development.
Sakamoto, Tachibana, and Takenaga \cite{Takenaga1999, Takenaga2000}
have pointed out 
that
breaking of the translation symmetry causes
breaking of the supersymmetry
because the supersymmetry includes the translation symmetry.
Hence the magnetic field may trigger supersymmetry breaking.
On the other hand, 
the MTG in an $ n $-torus is regarded as a generalization 
of the noncommutative torus, 
which attracts much attention recently in the string theory \cite{Connes}.
The $ B $-field in a compactified space naturally 
induces a noncommutative structure, which is described by the MTG.
Jackiw \cite{Jackiw} also showed that how the noncommutative structure
emerges in physical situations.
If we turn our attention to solid state physics,
we find another interesting application of the MTG also in this area.
Tranquada \cite{Tranquada} observed 
spontaneous formation of a charge density wave at a nonzero wave number
in a copper-oxide superconductor.
This ordered state is called the stripe phase, 
in which the translation symmetry is broken.
A similar stripe phase occurs commonly in a quantum Hall system \cite{Fradkin}.
Application of the MTG may help understanding of the stripe phases.

\section*{Acknowledgments}

I wish to thank 
Kin-ya Nakamura, whose thesis motivated this work.
I wish to thank 
M. Sakamoto for stimulating discussions.
This work originates from collaborations with him.
I thank A. Asada,
who taught me 
classification of $ S^1 $-fiber bundles by the first Chern class.
I appreciate valuable comments by
K. Fujii,
S. Higuchi,
S. Iida,
T. Iwai,
N. Maeda, 
N. Miyazaki, 
J. Nishimura,
Y. Ohnuki,
and
I. Tsutsui.
This work was partially supported by Grant-in-Aids for Scientific Research 
from
Ministry of Education, Culture, Sports, Science and Technology of Japan.

\baselineskip 4mm 
\vspace{12mm}
\baselineskip 6mm 
\section*{Note added to the proof}

After acceptance for publication of this article
we obtained more strong results on 
the magnetic translation group in $ n $ dimensions.
As concerns $ \gamma_i $ in (8.1) and $ \mu_i, \nu_i $ in (8.4),
we proved that $ \mu_i = \nu_i = 1/\gamma_i $.
Consequently, the definition (8.7) means simply that $ d_i = \nu_i $.
Eq. (8.29) is also simplified as $ c_i = \nu_i $.
In (8.31) the dimension of the irreducible representation becomes
$ \Pi_{i=1}^l \nu_i $.
{}Finally, the number of inequivalent representations (8.32) is reduced to one.
As a corollary,
we can show that 
$ \nu_1 = \nu_2 = d = c = 1/\gamma $
and hence 
$ \ell = d \gamma = 1 $ in Sec. 7.
More strongly, we can prove that
$ \nu_1 = \mbox{GCD} \, \{ b_1, b_2, b_3 \} $
for the three-dimensional magnetic field.
Proofs of these statements are to be published elsewhere.

\end{document}